\documentclass[twocolumn,showpacs,preprintnumbers,amsmath,amssymb]{revtex4}

\usepackage{graphicx}
\usepackage{dcolumn}
\usepackage{bm}
\usepackage{color}

\begin{document}


\setlength{\parindent}{1em}

\title{High-Accuracy Microwave Atomic Clock via Magic Optical Lattice
}

\author{Xiaoji Zhou}
\author{Xuzong Chen}
\author{Jingbiao Chen}\thanks{Electronic address: jbchen@pku.edu.cn }

\affiliation{School of Electronics Engineering $\&$ Computer
Science, Peking University,\\ Beijing 100871, P. R. China
}%


\date{\today}

\begin{abstract}
A microwave atomic clock scheme based on Rb and Cs atoms trapped in
optical lattice with magic wavelength for clock transition is
proposed. The ac Stark shift of clock transition due to trapping
laser can be canceled at some specific laser wavelengths. Comparing
with in fountain clock, the cavity related shifts, the collision
shift, and the Doppler effect are eliminated or suppressed
dramatically in atomic clock when the magic optical lattice is
exploited. By carefully analyzing various sources of clock
uncertainty, we conclude that a microwave atomic clock with an
accuracy of better than $2\times10^{-17}$ is feasible, which is of
the same accuracy as the expected best optical atomic clock.

\end{abstract}


\maketitle

After the realization of Zacharias' idea of atomic
fountain~\cite{Ramsey}, many advances in high-precision measurement
physics, including the test of fundamental constants
stability~\cite{Diddams,Bize}, have been achieved with atomic
fountain clock with a total fractional uncertainty of
$3.5\times10^{-16}$~\cite{Diddams,Bize,Heavner,Vian}. On the other
hand, many critical improvements of optical atomic
clock~\cite{Metrologia,Takamoto,Katori,Ludlow,Margolis} in the past
several years indicate that a potential competitor with much better
uncertainty than the microwave atomic clocks will soon be the new
generation of atomic clocks. For example, various experimental
schemes of optical clock based on alkaline-earth atoms and Yb atoms
with an uncertainty of better than $2\times10^{-17}$ are
proposed~\cite{Takamoto,Hong,Santra,chen}. Except the active optical
clock~\cite{chen}, all passive optical clock schemes use neutral
atoms trapped in optical lattices formed by ``magic" wavelength,
which can be used to adjust the ac Stark shift of clock transition
as fine as $1\times10^{-3}$Hz~\cite{Takamoto,Katori,Ludlow}.
Advantages of single ion clock and the conventional neutral atom
clock are combined there since millions of neutral atoms are trapped
within Lamb-Dick regime for long time clock-laser interrogation. The
most recent evaluation of atomic fountain clock is one order of
magnitude better than that of the best achieved optical
clock~\cite{Margolis,Metrologia}. However, it is obvious that, in
atomic fountain configuration with laser cooled Cs and Rb, to reach
a fractional uncertainty of microwave frequency standard to the
level of $1\times10^{-17}$ is a real technical challenge.
Difficulties mainly arise from the effects of second-order Zeeman
effect, collisions between the Cs atoms, microwave cavity, atomic
motion and blackbody radiation. For microwave clock transition, it
has been preliminarily precluded the possibility that a blue-detuned
dipole trap will supersede a fountain due to hundreds of mHz
residual ac Stark shift of clock transition in Na atom
experiment~\cite{Dav, Friedman}.

However, in this letter, we present the magic wavelengths for
$^{133}$Cs and $^{87}$Rb clock transitions. Our calculations show
that the ac Stark shift of Cs and Rb clock transitions can be
canceled at some specific magic wavelengths of trapping laser. Based
on a careful analysis of various sources of clock uncertainty
between fountain and lattice configurations, we conclude that a
microwave atomic clock based on Cs and Rb atoms trapped in magic
wavelength lattices is able to reach an accuracy of
$2\times10^{-17}$.

To realize such a microwave clock in optical lattices, it is
essential that the polarizabilities of the two clock states are
matched to high accuracy at the magic wavelength. The frequency
shift of clock transition induced by the axially symmetric electric
field $E_{z}$ with gradient $E_{zz}$ of the lattice laser is
described as~\cite{Fuentealba},
\begin{multline}
 \delta\nu=-\frac{1}{2}(\delta\alpha/h)
E_{z}^{2}-\frac{1}{4}(\delta C/h) E_{zz}^{2} \\
-\frac{1}{4}(\delta B/h) E_{z}^{2}E_{zz}-\frac{1}{24}(\delta
\gamma /h) E_{z}^{4}\ldots,
\end{multline}
where $h$ is the Plank constant, $\delta\alpha$, $\delta C$,
$\delta B$ and $\delta\gamma$ are the difference of the dynamic
dipole polarizabilities, quadrupole polarizabilities,
dipole-dipole-quadrupole hyperpolarizabilities and second dipole
hyperpolarizabilities between two clock states~\cite{Fuentealba}.

When $\delta\nu$ equals zero, we get the exact magic wavelength for
trapping laser. We first consider the dominant term, the first one
on the right hand of Eq.(1), which is determined by the differential
polarizability $\delta\alpha_{0}$ (both scalar and vector part) and
the tensor part $\delta\alpha_{2}$. The ac Stark shift of atomic
hyperfine state $|n,J,F,m_{F}\rangle$ for $\delta\alpha_{0}$
~\cite{Steck,Suter,Grimm,Degenhardt, Romalis} is,

\begin{multline}\label{eq:impor}
\delta \nu_{i}=-\frac{3\pi c^2 I_{L}}{2h}\sum_{i\neq j}
\frac{A_{ij}}{\omega_{ij}^{2}(\omega_{ij}-\omega)^{2}}(2J'+1)(2F'+1)\\(2F+1)
{\begin{pmatrix}F'&1&F\\m'_{F}&p&-m_{F}\end{pmatrix}}^{2}{\begin{Bmatrix}J&J'&1\\F'&F&I\end{Bmatrix}}^{2},
\end{multline}
where $I_{L}=(\varepsilon_{0}c/2)|E_{z}|^{2}$ is the intensity of
the trapping laser, $c$ the light velocity, $A_{kj}$ the Einstein
spontaneous coefficient, and the coefficients in round and curly
brackets are the $3J$ and $6J$ symbol, respectively. They describe
the selection rules and relative strengths of the transitions
depending on the angular momenta, their projections $m_{F}$, the
nuclear spin $I$ and the polarization $p=0,\pm1$ which stands for
$\pi$ and $\sigma^{\pm}$ transition of related states. The primed
quantum numbers refer to the excited states $n'P_{J'}$ coupled to
the ground state by allowed electric diploe transitions.We take
the needed transition frequencies $\omega_{ij}$ and $A_{ij}$
coefficients from the collected data of Kurucz and
Bell~\cite{Kurucz}.

The scalar light shift of clock transition for Cs irradiated by a
linearly polarized laser is shown in Fig.1. The ac Stark shift of Cs
clock transition due to $6P_{1/2}$ and $6P_{3/2}$ states are
negative at some wavelength region as the dotted lines show, but the
contribution from $7P$ as the dashed lines and higher states as the
dot-dashed line are positive. The combination of the above shifts
results in cancelation of ac Stark shift of Cs clock transition due
to the linearly polarized trapping laser at some magic wavelengths.
Our calculation predicts more than one magic wavelengths for Cs
clock transition at $402.2$nm, $427.0$nm, $493.0$nm, $604.8$nm as
showed in Fig.1 and for Rb at $370.4$nm, $395.6$nm, $452.3$nm,
$549.7$nm which are labeled with cross-mark in Fig.2. The tuning
rate $(2\pi d\nu_\text{clock} )/d\omega$ is as low as
$6\times10^{-14}$. The depth of lattice trap is $167$kHz ($8\mu$K)
for Cs with laser intensity of $10$kW/cm$^{2}$ at wavelength of
$604.8$nm, and the vibrational frequency is about $34$kHz.

\begin{figure} \centering
\includegraphics[height=5cm]{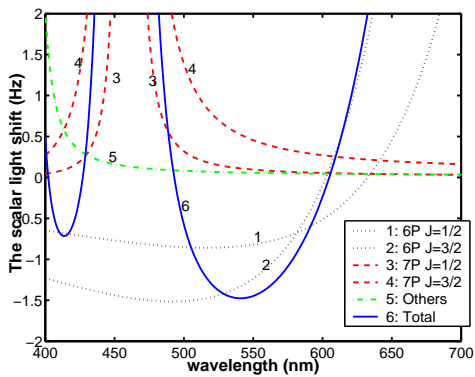}
\caption{The scalar light shift of Cs clock transition when atoms
irradiated by a linearly polarized laser with intensity of
10kW/cm$^{2}$.}
\end{figure}

\begin{figure} \centering
\includegraphics[height=5cm]{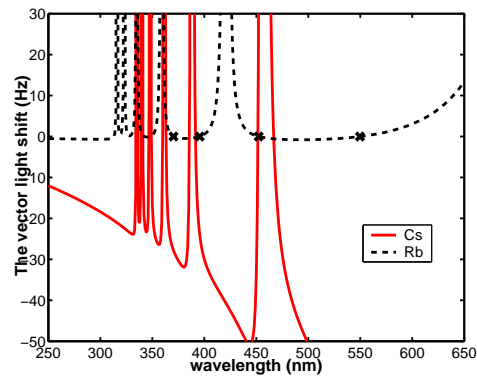}
\caption{The vector light shift of Cs and Rb clock transition when
atoms irradiated by $\sigma^{\pm}$ laser with intensity of
10kW/cm$^{2}$.}
\end{figure}

If atoms are irradiated by $\sigma^{\pm}$ laser, then $p=\pm 1$ in
Eq.(2), the consequent vector light shift of Cs and Rb clock
transitions are shown in Fig.2. For $^{87}$Rb atom, the ac Stark
shift due to $\sigma^{\pm}$ light at the magic wavelength of $\pi$
light is also very small, which is $2.3\times 10^{-3}$ Hz  at
549.7nm with a laser power of $10$kW/cm$^{2}$. The magic wavelength
of the vector light shift is 549.6nm, and its tuning rate $(2\pi
d\nu_\text{clock} )/d\omega$=$2.5\times10^{-14}$ are very close to
that of the scalar light shift as mentioned above. But the vector
light shift of Cs clock transition is much larger than that of $\pi$
light. Considering a real 3D blue-detuned optical lattice, the
trapped cold atom experiences not only the $\pi$ light but also
$\sigma^{\pm}$ light. Thus for Cs, the magic wavelength is mainly
determined by the vector light shift. Even though, the tuning rate
$(2\pi d\nu_\text{clock} )/d\omega$ is as low as $1.5\times10^{-12}$
at a magic wavelength of 471nm for $10$kW/cm$^{2}$ lattice laser
intensity with 3D laser polarization configuration of
$E_{x}(\hat{e}_{z}), E_{y}(\hat{e}_{z}), E_{z}(\hat{e}_{x})$, and is
good enough to control the light shift cancelation. It seems,
considering the polarization, $^{87}$Rb is a better candidate than
Cs to build a microwave atomic clock with atoms trapped in 3D
optical lattice. The following analysis is based on Cs, but is also
suitable for Rb.

The tensor light shift of clock transition $\delta\alpha_{2}$,
arising from the third-order perturbation when the hyperfine
interaction is taken into account~\cite{Ospelkaus}, is roughly
$\delta_{hfs} /\delta$ of the $\alpha$ term, where $\delta_{hfs}$
is the hyperfine splitting of the coupled excited state and
$\delta$ is the trapping laser detuning of that
state~\cite{Katori}. The tensor light shift of Cs clock transition
is of the order mHz shown in Fig.3. The tuning rate $(2\pi
d\nu_\text{clock} )/d\omega$ is as low as $1\times10^{-19}$, much
smaller than that of scalar light shift, hence it only modifies
the magic wavelength a little.

The static electric quadrupole polarizability $C/h$,
dipole-dipole-quadrupole hyperpolarizability $B/h$, the second
hyperpolarizability $\gamma/h$ of Cesium ground
state~\cite{Fuentealba} are $7.3\times10^{-25}$Hz$m^{4}/V^{2}$,
$4.65\times10^{-24}$Hz$ m^{4}/V^{3}$,
$1.05\times10^{-24}$Hz$m^{4}/V^{3}$ respectively. Under the static
field approximation, a $10$kW/cm$^{2}$ trapping laser causes
$5\times10^{-5}$Hz, $7.5\times10^{-6}$Hz and $2.5\times10^{-8}$Hz
shifts of clock transition corresponding to $\delta C$, $\delta B$,
and $\delta\gamma$, respectively. Atom is moving back and forth in
the laser intensity minima of 3D blue-detuned lattice
trap~\cite{Dav, Friedman} at the oscillating frequency of lattice,
this effect results in the cancelation of shift due to $\delta B$
term and negligible broadening due to $\delta C$ term. These shifts
are less than $5\times10^{-8}$Hz when the trapping laser power is
stabilized to $0.1\%$. These shifts can also be compensated by
scalar light shift easily. The ac Zeeman shift of clock transition
is only about $1\times 10^{-10}$Hz if the same laser power is
assumed.

\begin{figure} \centering
\includegraphics[height=5cm]{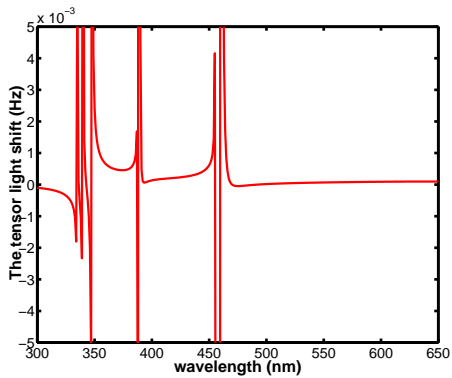}
\caption{The tensor light shift of Cs clock transition irradiated by
linearly polarized laser with intensity of 10kW/cm$^{2}$.}
\end{figure}

Fluctuations of the trapping laser intensity and frequency,
scattering of the trapping light, and atomic collisions with
background atoms, which are the three mechanisms for heating atoms,
mainly contribute to the trap lifetime. For the blue-detuned optical
lattice trap, the atoms are trapped at the minima of laser
intensity(the nodes of standing wave), the trap lifetime is larger
than $100s$, since the Rayleigh scattering rate is about
$0.01s^{-1}$ and the Raman scattering rate is as low as
$5\times10^{-6}s^{-1}$~\cite{Romalis}. For the background pressure
of the order of $1\times 10^{-11}$Torr, the trap time can be more
than 300s~\cite{Hara,Granade}. In the case when the lattice trapping
laser is stable, the heating arising from laser intensity noise and
beam pointing fluctuation~\cite{Hara,Granade,Gehm} is negligible
then the trap lifetime is finally limited only by Rayleigh
scattering rate. The microwave pulses of clock transition are
radiated by a dipole antenna close to the lattice
position~\cite{Dav, Kuhr}. The alternative method of clock
transition interrogation is performed by Raman transition with
microwave modulated laser.

We analyze the various sources of clock uncertainty with a
comparison between fountain and lattice configurations as follows
and list them in the Table I. In lattice clock configuration,
considering atoms are confined within a very small size like
$(100\mu$m)$^{3}$ for $10^{6}$ atoms confined in lattice with
lattice spacing of 300nm, the temperature can be easily controlled
to $¡À0.1K$ , thus the uncertainty of fractional blackbody shift is
$¡À1.3\times 10^{-17}$ or even better. Moreover, this $100\mu $m
level location of atoms position results in a much better stability
of gravitational redshift. But its uncertainty is determined only by
the calibration of orthometric height of atoms position. This has
potential applications in monitoring the change of gravitational
potential due to moon or tide in real-time when the short-term
stability is good. The spin-exchange collision shift in a Cs
fountain clock is at the level of $2\times 10^{-14}$, which can be
corrected to as low as $1\times10^{-16}$~\cite{Heavner} at low
density. In lattice clock configuration using 3D lattice with less
than unity occupation, the collision shift will be eliminated
completely~\cite{Katori}.

When a single microwave photon recoils during the Cs clock
transition, the relative frequency shift is
$\Delta\nu_\text{recoil}$ $=(\hbar k)^{2}/(2M\hbar \omega
)\approx1.4\times10^{-16}$, which was set as the uncertainty in a
fountain clock~\cite{Vian}. In lattice, this recoil shift is within
the spectral line width causing the residual 1st-order Doppler
effect as analyzed below. Assuming the direction of microwave vector
misalignment can be calibrated to $\leq$1mrad, the microwave recoil
shift can be calibrated to $<1\times10^{-18}$ in lattice clock.
Microwave power-dependent shift including distributed cavity phase
shift and microwave leakage shift is $
<2\times10^{-16}$~\cite{Heavner} and cavity pulling is about
$1\times10^{-16}$~\cite{Szymaniec}. The shift due to majorana
transition is expected less than $ 2\times10^{-17}$, which is set as
uncertainty~\cite{Heavner} in a fountain clock.While in the lattice
clock, there is no cavity pulling when using the traveling microwave
to interrogate the lattice-trapped atoms or via Raman process, and
no majorana transition when the residual magnetic field fluctuation
is much smaller than the C-field. Cavity phase shift~\cite{Li} and
microwave leakage shift can be eliminated since the Raman laser beam
can be extinguished by mechanical shutter.
\begin{table}
\caption{\label{tab:table 1}Comparison of uncertainties sources
between fountain clock and expected lattice clock. All values are
fractional frequency in $10^{-17}$.}
\begin{ruledtabular}
\begin{tabular}{lrr}
Clock configuration&Fountain\footnote{The reported minimal
uncertainties of fountain clock. Residual $1^{st}$-Doppler is
from~\cite{Bize}, the cavity pulling and the second-order Doppler
shift are from~\cite{Szymaniec}, microwave recoil is
from~\cite{Vian},
others are from~\cite{Heavner}.}&Lattice\\
\hline
  Light shift(ac Stark shift) & 0.001& $<$1, adjustable \\
  Blackbody shift & 26& 1.3 \\
  Gravitational redshift& 3 & 1 \\
  Collision shift & 10 & 0 \\
  cavity pulling & 1& 0 \\
  Microwave recoil& 14 & 0.1 \\
  Microwave leakage & 20 & 0 \\
  cavity distributed phase & 2& 0 \\
  Microwave spectral purity& 0.3 & 0.3 \\
  $2^{nd}$-order Zeeman & 2 & 0.1 \\
  Majorana transition & 2& 0.1 \\
  Residual $1^{st}$-Doppler & 2& 0 \\
  $2^{nd}$-order Doppler & 2& $10^{-4}$ \\
  Rabi and Ramsey pulling & 0.01& 0.1 \\
  Bloch-Siegert  & 0.01 & 0.01 \\
  Background gas collision & 0.1& 0.1 \\
  The dc Stark effect & 2& 0.1 \\
  TOTAL UNCERTAINTY & 38 & 2.0 \\
\end{tabular}
\end{ruledtabular}
\end{table}

For Cs atoms, the second-order Zeeman shift due to effect of static
magnetic field $B_{c}$ in the unit of Gauss on clock transition is
$427\times B_{c}^{2}$Hz. The C-field in a fountain clock is about
80nT=0.8mG, the uncertainty of fountain clock from C-field is
$2\times10^{-17}$~\cite{Heavner}. In the lattice clock
configuration, since the atoms are confined within a very small size
like ($100\mu$m)$^{3}$, the requirement for magnetic field shielding
and stability techniques will be dramatically relaxed, then the
clock uncertainty from C-field $B_{c}$ should be controlled less
than $1\times 10^{-17}$ without any difficulty more than that of
fountain clock.

This residual first order Doppler shift is from the product of two
vectors $\vec{k}\cdot\vec{v}$, where $\overrightarrow{k}$ is the
wave vector of microwave, $\overrightarrow{v}$ is the atomic
velocity. Assuming the direction misalignment is $\leq$1mrad,
$\upsilon=3$m/s, for a traveling wave, these parameters result in a
frequency shift of 0.02Hz. In a fountain, when the clock microwave
signal is fed into the cavity ``from the left'' or ``from the
right'', the observed shift is about
$2.5\times10^{-5}$Hz~\cite{Bize}, this means the residual traveling
wave component in the fountain cavity is near $10^{-3}$ of the
standing wave component. In a lattice clock, assuming with the same
microwave intensity(the same pulse time about $10$ms), the microwave
is a pure traveling wave, thus it is 1000 times of the traveling
wave in fountain cavity. And, the atomic average velocity during the
microwave pulse period time 10ms is $\langle
v\rangle=0.2\times300$nm$/10$ms$=6\mu $m/s, where 300nm is the
lattice spacing, 0.2 means the atom is confined within the $20\%$
central range near the center of a lattice trap, and 10ms is the
microwave pulse period. Considering the atomic motion within 3D
lattice minima with a oscillating frequency during the microwave
pulse period time 10ms, the net effect of atomic ensemble is a
Doppler spectral-broadening of $\pm2\times10^{-4}$Hz. However, there
is no net residual first-order Doppler shift in a lattice clock. The
second-order Doppler shift due to relativistic time dilute is
$1\times 10^{-17}$~\cite{Szymaniec} in a fountain clock, and will be
$<1\times 10^{-21}$ in lattice clock configuration. Rabi pulling,
Ramsey pulling, microwave spectral purity, and Bloch-Siegert shift
do not depend on the clock configuration, but they are determined by
the C-field, the performance of synthesizer and other parameters for
both clocks.

It is possible to reach $N=10^{6}$ of cold atoms for a lattice
clock, therefore the stability can be improved to
$\sigma_{y}(\tau)=5\times10^{-15}/\sqrt{\tau}$ assuming
interrogation time is 20s.

In summary, we have discussed the feasibility of a microwave atomic
clock based on Rb and Cs atoms trapped in optical lattice with magic
wavelength. By a careful comparison of various sources of clock
uncertainty between fountain and lattice configurations, we conclude
that a microwave atomic clock with an accuracy of $2\times10^{-17}$
is feasible, which is of the same precision as the expected optical
atomic clock. This clock scheme overcomes the formidable hurdles in
fountain clock arising from the second-order Zeeman effect,
collisions between the Cs atoms, the microwave cavity, the atomic
motion and the blackbody radiation. Besides the potential
application in high-precision measurements, our clock scheme
promises that microwave based second definition in SI units can be
realized one order of magnitude more precisely. Moreover, the
practical size of microwave clock can be reduced to only one third
of current fountain size or even smaller, and the idea in this
letter is also applicable for designing primary atomic clock for
international space station~\cite{Bize, ISS}.

We acknowledge helpful discussions with Yiqiu Wang and Ruoxin Li. We
also thank Mei Zhang for her critical reading. This work is
supported by MOST under Grant No.2005CB3724500, and NSFC under
Grants 60178016, 10104002, 10574005.\\

\end{document}